\documentclass{article}
\usepackage{spconf,amsmath,graphicx,color}
\usepackage{amsfonts}
\usepackage{subcaption}
\usepackage{multirow}
\usepackage{booktabs}

\title{Multi-Domain Adaptation by Self-Supervised Learning for Speaker Verification}
\name{Wan Lin$^{1,3}$, Lantian Li$^{2*}$, Dong Wang$^{1*}$}
\address{$^1$Center for Speech and Language Technologies, Tsinghua University, China \\
         $^2$Beijing University of Posts and Telecommunications, China  \\
         $^3$Shen Zhen University, China}
\begin{document}
\ninept
\maketitle
\begin{abstract}
In real-world applications, speaker recognition models often face various domain-mismatch challenges, 
leading to a significant drop in performance. Although numerous domain adaptation techniques have been developed to address this issue, 
almost all present methods focus on a simple configuration where the model is trained in one domain and deployed in another. 
However, real-world environments are often complex and may contain multiple domains, making the methods designed for one-to-one adaptation suboptimal. 
In our paper, we propose a self-supervised learning method to tackle this multi-domain adaptation problem. 
Building upon the basic self-supervised adaptation algorithm, we designed three strategies to make it suitable for 
multi-domain adaptation: an in-domain negative sampling strategy, a MoCo-like memory bank scheme, and a CORAL-like distribution alignment. 
We conducted experiments using VoxCeleb2 as the source domain dataset and CN-Celeb1 as the target multi-domain dataset. 
Our results demonstrate that our method clearly outperforms the basic self-supervised adaptation method, 
which simply treats the data of CN-Celeb1 as a single domain. 
Importantly, the improvement is consistent in nearly all in-domain tests and cross-domain tests, demonstrating the effectiveness of our proposed method.
\end{abstract}

\section{Introduction}

Speaker verification aims to determine whether a given speech belongs to a particular speaker. In recent years, the development of deep neural networks (DNNs) has significantly improved the performance of speaker verification systems~\cite{hanifa2021review}. The deep embedding architecture, in particular x-vector, has been mostly widely adopted~\cite{snyder2018x}. It involves a deep embedding model to extract speak vectors for enroll/test utterances and a simple scoring back-end based on cosine distance.
Recently, the deep embedding approach has achieved state-of-the-art performance in numerous benchmark evaluation tasks~\cite{sadjadi20222021,huh2023voxsrc}.

In spite of the brilliant success, the true performance of speaker verification systems
in real deployment is still unsatisfactory. An important reason is that the acoustic condition
in real applications often substantially deviates from that of the training data. 
Unfortunately, retraining a model is costly, so how to adapt a well-trained model to
meet the deployment condition becomes crucial.

A straightforward approach is to fine-tune an existing model using data from the
target domain. If the data is abundant, the model can be well adapted to
the target domain.
However, collecting and labeling target-domain data is costly or even impractical.
To tackle the problem,
Chen et al.~\cite{chen2021self} presented a self-supervised learning approach
based on contrastive loss, which does not require precise speaker labels
for the target data, but only positive and negative pairs that can be 
easily produced. Recently, Mao et al.~\cite{mao2023cluster} employed the same approach
to train a speaker embedding model using the target domain data, and then
combined this model with k-mean clustering to produce pseudo labels for unlabelled
data.

The self-supervised adaptation is highly promising. 
However, the present work mainly focuses on adapting the model from one domain (source) to another domain (target).
This approach does not match the requirements of many real applications that involve processing data from multiple domains.
Consequently, it is also necessary to consider the adaptation to multiple domains.
And this problem, i.e., \emph{multi-domain adaptation}, has not been well studied yet.
Most research just treats all the 
target data as from a single domain, even if they are in fact 
from multiple domains~\cite{chen2021self,mao2023cluster}.

In this paper, we present a multi-domain adaptation approach based on self-supervised learning (SSL). 
The contrastive self-supervised model~\cite{chen2021self} that treats all the target data as from a single domain 
is used as the baseline, which we name \emph{single-domain SSL adaptation}. 
Three strategies are designed to make the multi-domain SSL adaptation feasible:
(1) An in-domain negative sampling, that enforces negative samples
being sampled from the same domain as the pivot speech. This avoids mistaking inter-domain variation
for inter-speaker variation; (2) A MoCo-like memory bank that caches recently seen samples, 
to address the negative impact of in-domain negative sampling; (3)
CORAL-like distribution alignment, which aligns the distributions of speaker vectors from different domains.

Our experiments were conducted using a multi-genre speaker dataset CN-Celeb1~\cite{fan2020cn} to settle on a multi-genre speaker verification task.
For clarity, we treat each genre as a particular domain in our experiments. 
The VoxCeleb2 dataset~\cite{nagrani2020voxceleb} was used to train an initial model,
and the CN-Celeb1 dataset was used to perform multi-domain adaptation. 
Our multi-domain SSL adaptation approach achieved an EER of 11.54\% on the CN-Celeb1 evaluation set, 
resulting in a 9.35\% relative improvement compared to the single-domain SSL adaptation baseline. 
Moreover, we find that the improvement is consistent over 
all the in-domain and cross-domain test scenarios, demonstrating the stability of the proposed method.


\begin{figure*}[htbp]
  \centering
   \includegraphics[width=0.94\linewidth]{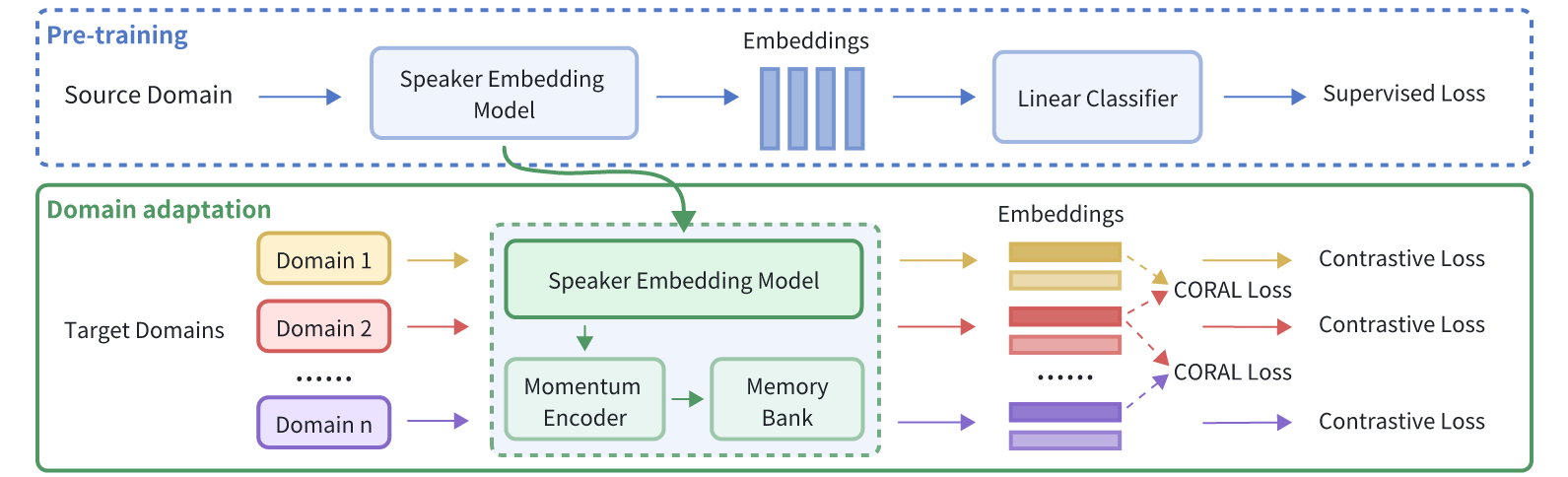}
   \caption{The framework of our proposed multi-domain adaptation by contrastive self-supervised learning.}
  \label{fig:framework}
\end{figure*}

\section{Related work}

Domain mismatch is a general problem and hot topic in speaker recognition. 
Early studies focused on adapting the back-end scoring model, e.g., ~\cite{McCree2014UnsupervisedCA},
and recent approaches more focus on adapting the embedding model~\cite{li2022editnet, hu2022class, chen2021self, wang2021adversarial, hu2022domain}.
One approach is to minimize the discrepancy between the distributions of speaker vectors in the source and target domains~\cite{lee2019coral+, li2022coral++, hu2022class, lin2018reducing, lin2020multi}. Representative algorithms include Correlation Alignment (CORAL)~\cite{sun2016deep} and Maximum Mean Discrepancy (MMD)~\cite{lin2018reducing}, WBDA (within-between distribution alignment)~\cite{hu2022class}.
The second approach employs
adversarial training to make the speaker vectors domain-independent~\cite{wang2021adversarial, rohdin2019speaker, bhattacharya2019generative, wei2023contrastive}. All these approaches need both the source and target data, thus not suitable for real-life applications. 
However, the idea of distribution alignment can be used in multi-domain adaptation and has been integrated into
our multi-domain SSL adaptation approach. 



Contrastive self-supervised learning~\cite{zhang2021contrastive, lepage2022label, ravanelli2018learning, chen2020simple, he2020momentum} has been 
widely employed in speaker recognition. 
We employ this approach to perform domain adaptation. 
While this approach has been explored previously in~\cite{chen2021self}, our approach differs in that we specifically consider multi-domain information, whereas they treat all the data from a single domain.

\section{Method}

In this section, we first review the contrastive self-supervised learning framework. Then, we present the three ingredients in our multi-domain SSL adaptation approach, as shown in Fig.\ref{fig:framework}.

\subsection{Contrastive self-supervised learning}

Contrastive self-supervised learning samples positive and negative pairs, and train the model to 
pull positive pairs closer and push negative pairs further apart.
In speaker recognition, two segments from the same utterance can be regarded as positive pairs, and two segments from different 
utterances are negative pairs.

Formally, suppose a mini-batch consisting of $N$ utterances, and each utterance is denoted by $u_i$.
Two non-overlapping segments are randomly sampled from the same utterance $u_i$, denoted by $u_{i,1}$ and $u_{i,2}$ respectively.
After data augmentation, $u_{i,1}$ and $u_{i,2}$ are fed into the speaker embedding model $f_{\theta}$ 
to produce speaker vectors $e_{i,1}$ and $e_{i,2}$. All the utterances in the mini-batch are processed in the same way, and 
then a contrastive loss~\cite{oord2018representation} is computed as follows:

\begin{equation}
\mathcal{L}_{\mathrm{CL}}=\frac{1}{N} \displaystyle\sum_{i=1}^{N}{\frac{\mathrm{sim}(e_{i,1},e_{i,2})}{\textstyle\sum_{j=1}^{N}{\mathrm{sim}(e_{i,1},e_{j,2})}}},
\end{equation}

\noindent where $i$ and $j$ index utterances, and $\mathrm{sim}(\cdot,\cdot)$ measures distance between two speaker vectors, defined as follows:

\begin{equation}
\label{eq:sim}
\mathrm{sim}(x,y)=\exp(\cos(x,y)/\tau),
\end{equation}
\noindent where $\tau$ is a temperature parameter.

\subsection{In-domain negative sampling}
\label{sec:DSNS}

Traditional contrastive self-supervised learning samples negative pairs without any constraint. 
This is questionable when the adaptation data involves multiple domains, as utterances from 
two different domains represent both inter-speaker variation and inter-domain variation (suppose they are 
truly from two speakers). This leads to a biased estimation of the inter-speaker covariance. 

To address this issue, we propose an in-domain negative sampling strategy:
Firstly, we partition the samples within a mini-batch according to their domain labels, 
and then create negative pairs by sampling two utterances that can exclusively originate from the same domain.

\subsection{MoCo-like memory bank}

An issue of the in-domain negative sampling is that the number of negative pairs 
could be insufficient. 
We borrowed a training strategy from the MoCo (Momentum Contrast) approach~\cite{he2020momentum}.
MoCo designs a `memory bank' to store embeddings from recent mini-batches.
All these embeddings in this memory bank can be used as negative samples, thereby increasing the diversity of negative pairs.
In our approach, a momentum encoder $f_{\theta_k}$ is introduced, whose parameters are 
from the speaker embedding and are updated by a momentum update strategy:

\begin{equation}
\theta_k=m\theta_k+(1-m)\theta,
\end{equation}

\noindent where $m$ is a momentum coefficient, which is set to 0.999 in our experiments. 
This momentum encoder accepts samples in the previous mini-batches and produces the memory bank.

\subsection{CORAL-like distribution alignment}

Inspired by the success of DeepCORAL~\cite{sun2016deep} in domain adaptation, we apply the idea of distribution alignment to our multi-domain SSL adaptation approach.
Specifically, a multi-domain CORAL loss is introduced to regulate the adaptation:

\begin{small}
\begin{equation}
\mathcal{L}_{\mathrm{CORAL}}=\frac{2}{N(N-1)} \frac{1}{4d^2} \displaystyle\sum_{i=1}^{N} \sum_{j=i+1}^{N} \Vert \varSigma_{i}-\varSigma_{j} \Vert_F^2,
\end{equation}
\end{small}

\noindent where $N$ is the number of domains in the mini-batch, $\varSigma_{i}$ and $\varSigma_{j}$ are covariance matrices of domain $i$ and $j$ respectively.
$\Vert \cdot \Vert_F^2$ denotes the Frobenius norm. 

~~

The model is then adapted by minimizing the following loss:

\begin{equation}
\mathcal{L}=\mathcal{L}_{\mathrm{CL}}+\lambda \mathcal{L}_{\mathrm{CORAL}},
\end{equation}

\noindent where $\lambda$ is a hyper-parameter to balance the contrastive loss and the CORAL loss.

\section{Experiments}

\subsection{Data}

Two datasets were used in our experiments:
VoxCeleb2~\cite{nagrani2020voxceleb} as the source single-domain dataset and CN-Celeb1~\cite{fan2020cn} as the target multi-domain dataset.

\emph{VoxCeleb2}: A large-scale speaker dataset collected by the University of Oxford, UK.
In our experiments, the development set of VoxCeleb2 (Vox2.dev) was used to train the x-vector models, encompassing a total of 5,994 speakers.
Since the dataset was collected from a single media source YouTube and predominantly came from the interview scenarios,
we treat it as the \emph{source single-domain} dataset.
In addition, data augmentation was applied to enhance model robustness,
with the MUSAN corpus~\cite{snyder2015musan} used to generate noisy utterances and the RIRS corpus~\cite{ko2017study} used to create reverberant utterances.

\emph{CN-Celeb1}: A multi-genre speaker dataset collected by Tsinghua University, China.
It contains about 125k utterances from 997 Chinese celebrities.
The dataset covers 11 diverse genres, each of which can be regarded as a specific domain.
Therefore, we designate it as the \emph{target multi-domain} dataset. The entire dataset was split into two parts:
CNC1.dev, which contains 797 speakers, was employed for multi-domain adaptation training;
CNC1.eval, which consists of 200 speakers, was used for performance evaluation.
During multi-domain adaptation training, short utterance combination~\cite{chen2021self} was firstly employed to splice short utterances from the same speaker within the same genre
into longer utterances that are at least 5s in length. Subsequently, data augmentation was applied, akin to the one used in Vox2.dev.

\subsection{Settings}

We adopted the x-vector architecture to construct the speaker embedding model.
This model accepts 80-dimensional Fbanks as input features. An ECAPA-TDNN~\cite{desplanques2020ecapa} 
backbone then produces frame-level features, and the features are aggregated by an attentive statistic pooling and 
forwarded to a full-connection layer to produce utterance-level x-vectors.
The cosine distance of the corresponding x-vectors is used to score the similarity between two utterances in speaker characteristics.
In our experiments, the dimensionality of the x-vector was set to 192.
The additive angular margin softmax (AAM-Softmax)~\cite{xiang2019margin} was used to discriminate speakers with a margin value of 0.2 and a scale factor of 30.

In the target multi-domain adaptation phase, the temperature $\gamma$ in Eq.(\ref{eq:sim}) was set to 0.07.
For the memory bank, the size was set at 8,192.
For simplicity, we assigned equal weights to both the contrastive loss and the CORAL loss.

\begin{figure*}[h]
 \centering
 \includegraphics[width=0.92\linewidth]{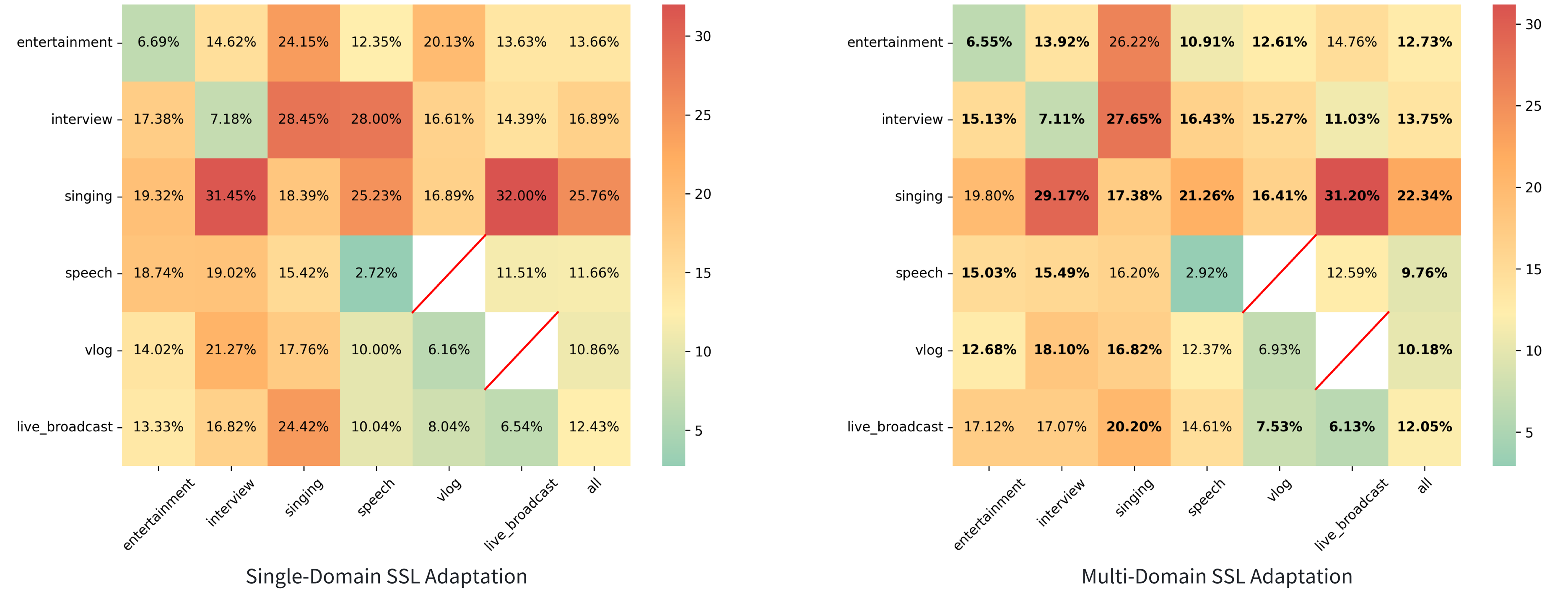}
 \caption{Multi-domain tests with two SSL adaptation methods.}
 \label{fig:cross}
\end{figure*}

\subsection{Basic Results}

We first present the basic results evaluated on CNC1.eval with different methods.
The results in terms of equal error rate (EER) and minimum of the normalized detection cost function (minDCF) are reported in Table~\ref{tab:basic}. Note that we used $C_{miss}$ = $C_{fa}$ = 1 and $P_{tar}$ = 0.05 in the cost function.

\begin{table}[!htbp]
 \caption{Results on CNC1.eval with different methods. `SL' denotes supervised learning, `FT' denotes fine-tuning, and `SSL' denotes self-supervised learning, `SD' and `MD' represent `single domain' and `multi-domain' respectively. }
 \centering
 \scalebox{0.82}{
 \begin{tabular}{clllcc}
  \toprule[1pt]
         No.    & Train Set   & Dev Set      & Mode           & EER(\%)   & minDCF            \\
   \midrule[1pt]
          1     & Vox2.dev    & -            & SL             & 14.22     & 0.5137                 \\
          2     & CNC1.dev    & -            & SL             & 12.24     & 0.5069                 \\
          3     & Vox2.dev    & CNC1.dev     & SL-FT          & \textbf{9.50}  &   \textbf{0.3991} \\
   \midrule[1pt]
          4     & CNC1.dev    & -            & SSL-SD      & 15.66     & 0.5822                 \\
          5     & Vox2.dev    & CNC1.dev     & SSL-SD      & 12.73     & 0.4812                 \\
          6     & Vox2.dev    & CNC1.dev     & SSL-MD (ours)      & \textbf{11.54} & \textbf{0.4551}   \\
   \bottomrule[1pt]
  \end{tabular}
  \label{tab:basic}}
\end{table}

Firstly, when comparing System 4 to System 2 and System 5/6 to System 3,
it can be observed that there is still a performance gap between SSL methods and SL methods.
This is understandable since SSL does not utilize any speaker information during training.

Secondly, it can be seen that System 5 outperforms System 1 and even comes close to System 2,
indicating the effectiveness of the basic single-domain SSL adaptation method in multi-domain adaptation.
This is consistent with the result reported in~\cite{chen2021self}.

Thirdly, the performance of our proposed multi-domain SSL adaptation method (System 6) notably 
surpasses that of the single-domain SSL adaptation method (System 5) and even outperforms the SL model in System 2.
This underscores the significance of domain information in multi-domain SSL adaptation and
also demonstrates that our multi-domain SSL adaptation method 
effectively leverages this information to achieve better performance.

\subsection{Multi-domain tests}

To provide a more detailed comparison between the single-domain SSL and our proposed multi-domain SSL methods, we construct multi-domain tests presented by a genre-to-genre performance matrix that encompasses both in-domain and cross-domain tests.
To mitigate the statistical bias in EER, we first selected 6 genres with the largest number of speakers and utterances.
Then we splice the utterances of each speaker in each genre to create 20-second enrollment segments, with the remaining utterances reserved for testing.
Using these enrollment and test utterances, we constructed in-domain and cross-domain test trials.

Fig.~\ref{fig:cross} shows the in-domain (diagonal) and cross-domain (off-diagonal) performance with the single-domain SSL (left) and multi-domain SSL (right) adaptation methods.
For each sub-figure, the numerical values shown in the blocks are the EER results under the enrollment genre corresponding to its row
and the test genre corresponding to its column,
and the last column shows the overall results that the enrollment is based on one genre and the test is on all the genres.
It is evident that in nearly all test cases (highlighted in bold in the right plot), our multi-domain SSL adaptation method consistently outperforms the single-domain SSL adaptation method,
providing further evidence of the effectiveness of our approach.

\subsection{Visualization}

\begin{figure}[!htp]
  \centering
  \includegraphics[width=1\linewidth]{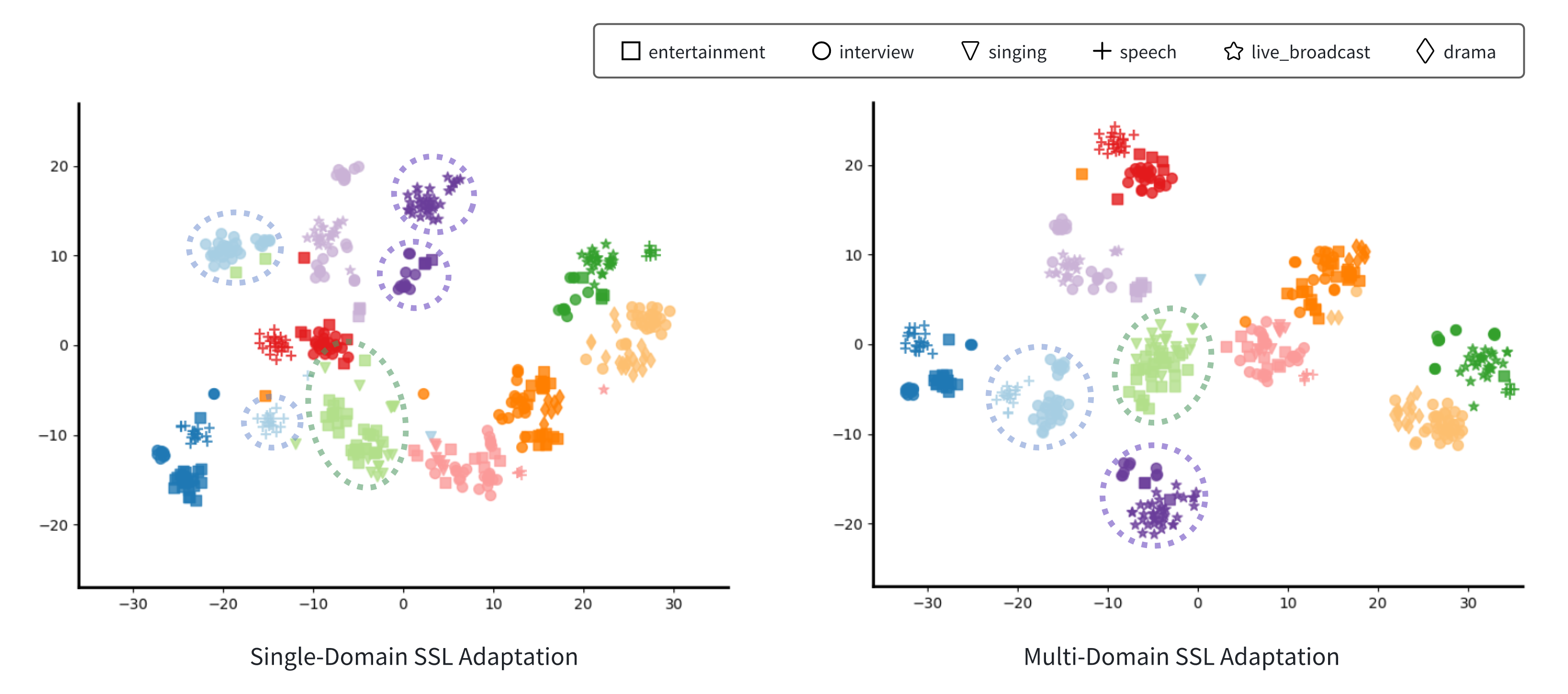}
  \caption{Multi-genre speaker distributions with single-domain SSL adaptation (left) and multi-domain SSL adaptation (right). The plots were produced by t-SNE~\cite{van2008visualizing},
  where each color represents a speaker and each shape represents a genre.}
  \label{fig:tsne}
\end{figure}

In this section, we analyze the single-domain and multi-domain SSL adaptation methods by visualization.
We randomly selected 10 speakers from the CNC1.eval dataset, each of whom covers more than 2 genres, and then sampled 50 utterances from each of these speakers.
The t-SNE toolkit~\cite{van2008visualizing} is applied to project the speaker x-vectors to a 2-dimensional space.
Fig.~\ref{fig:tsne} presents the results. It can be seen that our multi-domain SSL approach produces speaker vectors with lower intra-speaker variation and 
higher inter-speaker discrimination, as exemplified by the three speakers enclosed within the circles.
This result once again demonstrates the superiority of multi-domain SSL adaptation over single-domain SSL adaptation.

\subsection{Ablation study}

In this section, an ablation study is conducted to investigate the roles of various components involved in the multi-domain SSL adaptation method. Table~\ref{tab:ablation} shows the results.

\begin{table}[!htbp]
 \caption{Ablation study on different components in the multi-domain SSL adaptation approach. `IDNS' denotes in-domain negative sampling.}
 \centering
 \scalebox{0.97}{
 \begin{tabular}{llll}
  \toprule[1pt]
        No.  & Method                  & EER(\%)   & minDCF       \\
  \midrule[1pt]
        1    & Single-domain SSL        & 12.73     & 0.4812      \\
        2    & \quad + MoCo             & 12.44     & 0.4566      \\
  \midrule[1pt]
        3    & \quad + IDNS             & 12.66     & 0.5039      \\
        4    & \quad\quad + MoCo        & 11.69     & 0.4533      \\
        5    & \quad\quad\quad + CORAL  & 11.54     & 0.4551      \\
   \bottomrule[1pt]
  \end{tabular}}
  \label{tab:ablation}
\end{table}

Firstly, by comparing System 3 to System 1, it can be observed that the in-domain negative sampling is effective, but the performance gain is limited.
However, when the MoCo-like memory bank is involved, the performance is significantly improved (System 4 vs. System 3), demonstrating that the simple 
in-domain negative sampling suffers from insufficient negative samples, and the MoCo-like memory bank can offer compensation. In fact, the MoCo-like memory bank itself
also works (System 2 vs. System 1), though combining it with in-domain negative sampling leads to much better performance. 
Finally, by introducing the CORAL loss (System 5), we reached the multi-domain SSL that achieved the best results.

\section{Conclusion}

This paper proposed a multi-domain self-supervised learning (SSL) adaptation framework 
that adapts a pre-trained speaker model to make it match multiple domains.
The framework consists of three elaborate designs:
(1) An in-domain negative sampling strategy to eliminate the impact of inter-domain variation; (2) a MoCo-like memory bank scheme to
compensate for the negative impact of the in-domain negative sampling; (3) a CORAL-like distribution alignment to alleviate distribution shift across different domains.
Experiments were conducted using VoxCeleb2 as the source single-domain dataset and CN-Celeb1 as the target multi-domain dataset. 
The results demonstrated that our proposed multi-domain 
SSL adaptation method outperforms the basic single-domain SSL adaptation method and even 
surpasses the supervised multi-domain training.
Furthermore, a more comprehensive study showed that our method consistently improves performance 
in almost all multi-domain tests, confirming the stability and reliability of 
the proposed method. Future work will investigate the combination of the SSL adaptation method with other 
domain-invariant training methods such as adversarial training, or other domain alignment methods, 
e.g., within-class and between-class distribution alignment.

\newpage
\bibliographystyle{IEEEbib}
\bibliography{refs}

\end{document}